\documentclass[11pt,twocolumn]{article}
\usepackage[margin=1in]{geometry}
\usepackage{authblk}
\usepackage{amsmath,amssymb}
\usepackage{graphicx}
\usepackage{bm}
\usepackage{xcolor}
\usepackage{microtype}
\usepackage{abstract}
\usepackage{cite}

\newcommand{\vrf}[1]{\textcolor{magenta}{ #1}}

\title{A New Twist on Astigmatic Transformation: Beyond Laguerre–Gaussian Beams}

\author[1,$\dagger$]{Gabriela Flores-Cova}
\author[2,$\dagger$]{Bikash Kumar Das}
\author[1]{Carmelo Rosales-Guzman}
\author[3]{Valeria-Rodriguez-Fajardo}
\author[4,*]{Camilo Granados}

\affil[1]{Centro de Investigaciones en Óptica, A.C., Loma del Bosque 115, Colonia Lomas del Campestre, 371507 León, Gto., Mexico}
\affil[2]{Theoretisch-Physikalisches Institut, Friedrich-Schiller-Universität Jena, Max-Wien-Platz 1, D-07743 Jena, Germany}
\affil[3]{Departamento de Física, Universidad Nacional de Colombia Sede Bogotá, Carrera 30 No. 45-03, Bogotá, 111321, Colombia}
\affil[4]{Eastern Institute of Technology, Ningbo 315200, China}
\affil[*]{camilo.granados@eitech.edu.cn}
\affil[$\dagger$]{These authors contributed equally to this work}

\begin{document}

\twocolumn[
\maketitle

\begin{abstract}
Astigmatic transformation is widely used to convert the azimuthal phase structure of Laguerre--Gaussian vortex beams into intensity patterns from which the topological charge can be read out. Here, we show that this phase-to-intensity conversion is not just restricted to the Laguerre--Gaussian radial profile. We derive analytical expressions for the astigmatically transformed high-order Bessel--Gauss and perfect optical vortex fields, and find that, despite their substantially different radial structures, the number of intensity minima is directly set by the accumulated azimuthal phase winding. Numerical simulations and experimental measurements on both beam families confirm this result, establishing astigmatic transformation as a direct probe of the topological charge of scalar structured optical fields. We further extend the approach to non-separable vector Bessel--Gauss beams, in which the spatial and polarization degrees of freedom are coupled, and show that polarization-resolved detection allows the topological charge of the constituent spatial modes to be determined independently.
\end{abstract}
\vspace{1cm}
]

\section{Introduction}
Coherent light beams carrying orbital angular momentum (OAM), also known as optical vortex or phase-singular beams, possess helical wavefronts with a central phase singularity~\cite{Allen1992}. These beams are characterized by a discrete topological charge (TC), which defines the number of $2\pi$ phase shifts accumulated along the azimuth. Common examples include Laguerre--Gaussian (LG)\cite{Allen1992,Padgett:96}, Bessel--Gauss (BG)\cite{Khazaei:26,ArltDholakia2000}, and perfect optical vortex (POV) beams~\cite{Ostrovsky2013,Vaity2015,Ma2017,GarciaGarcia2014}. \vrf{}

The use of SLMs to generate vortex beams for driving nonlinear, non-perturbative light--matter interactions is fundamentally limited by their damage threshold, since these processes typically require high-intensity laser fields. In particular, in solid-state high-harmonic generation, the wavelength is also a constraint, since it is in the mid-infrared spectral regime. Consequently, developing techniques capable of accurately characterizing ultra-intense vortex beams is essential for enabling strong-field applications such as in high-harmonic generation.

While the radius of the maximum-intensity ring and the central dark core increase monotonically with TC for LG and BG beams, the radial intensity profile of POV beams is essentially independent of TC~\cite{Pinnell_19,Bikash:2026}.

Vortex beams can be broadly classified as scalar or vector depending on their polarization structure. Scalar vortex beams feature a uniform polarization state across their cross-section, whereas vector vortex beams (VVBs) exhibit spatially varying polarization distributions~\cite{Yan:2025}, making their spatial and polarization degrees of freedom non-separable~\cite{Gori1987,rosales2017simultaneous}. In both cases, the TC characterizes the underlying vortex structure, and its determination is essential for various applications~\cite{Baltrukonis2021,yu2016super,macdonald2002}.

Consequently, various methods have been developed to measure the TC, including interferometric approaches~\cite{BerkhoutBeijersbergen2008,Denisenko2009} and diffraction-based techniques that convert the vortex phase structure into a measurable intensity distribution~\cite{Liu2011,Hickmann2010,Fu2015,YONGXIN:2011,HanZhao2011,VAITY:2013,Chen:17}. Widely used configurations rely on axicons, tilted convex lenses, or angular double slits~\cite{HanZhao2011,VAITY:2013,Fu2015}, with the latter also applied to Bessel beams possessing fractional TCs~\cite{Zhu2018}. Other approaches exploit phenomena such as the rotational Doppler effect and the Talbot effect~\cite{Cheng:2022,Panthong:18}.

Astigmatic transformation (AT) using a cylindrical lens provides a simple yet powerful method for TC determination~\cite{Prabhakar2011,Kotlyar2016,Kotlyar2017,Peng2015}. For cylindrically symmetric vortex beams, the cylindrical lens transforms the azimuthal phase structure into a characteristic intensity pattern. In LG beams, the transformed intensity resembles a tilted Hermite--Gaussian (HG) mode, where the number of intensity minima between bright lobes yields the TC magnitude, and their orientation indicates its sign~\cite{fathollazade2025laguerre,fang2017examining}. Quantitative OAM measurements based on intensity moments at the cylindrical lens focus have been demonstrated for integer and fractional TCs~\cite{Alperin2016}, and AT has been applied to Bessel-like beams~\cite{Yuan2022} as well as harmonic radiation across different spectral ranges~\cite{Bikash:2026,GranadosOAM,GranadosPOV}. However, analytical expressions for astigmatically transformed fields are available only for specific cases, and how the radial field distribution determines the resulting intensity minima remains unestablished. This gap is central, as LG-based interpretations rely on a radial structure that differs fundamentally from the multi-ring profile of high-order BG beams and the TC-independent profile of POV beams.

Here, we investigate the AT of high-order BG, POV, and vector BG beams to establish how their azimuthal phase structure and distinct radial profiles shape the transformed intensity. For scalar BG and POV fields, we derive analytical expressions for the transformed distributions and show that the TC-dependent intensity minima persist despite their contrasting radial characteristics—namely, the oscillatory multi-ring profile of BG beams and the TC-independent ring radius of POV beams. Furthermore, we extend this approach to non-separable vector Bessel beams, demonstrating that polarization-projective measurements decouple the transformed intensity distributions of constituent spatial modes, enabling their respective TCs to be determined independently. Together, these results demonstrate that AT-based TC determination is robust across diverse radial field distributions and non-separable vector fields.

\section{Theoretical framework}
\subsection{A primer on astigmatic transformation by a cylindrical lens}

Astigmatic transformation provides a convenient means of converting the azimuthal phase structure of an optical vortex into a characteristic intensity distribution~\cite{Prabhakar2011,Kotlyar2016,Kotlyar2017,Peng2015}. A cylindrical lens introduces optical power only along one transverse direction and, in its focal plane, performs an one-dimensional (1D) Fourier transformation on the incident field along that direction. Consequently, the azimuthal phase structure of a vortex beam is mapped onto a transverse intensity distribution. For conventional vortex beams, this transformation produces characteristic intensity patterns from which the TC can be determined without the need for an external reference beam~\cite{Kotlyar2016,Kotlyar2017}.

Mathematically, the AT can therefore be represented by a 1D Fourier transform of the complex field amplitude along one transverse coordinate, while the orthogonal coordinate remains unchanged. Assuming that the transformation is performed along the $y$-axis, the transformed field can be written as:
\begin{eqnarray}
\label{eq:CL} 
\tilde{E}_{VB}(x,k_y) &=& \int_{-\infty}^{\infty} E_{VB}(x,y) e^{-ik_y y} dy 
\end{eqnarray}

where $E_{VB}(x,y)$ is the complex field amplitude of the incident vortex beam and $\tilde{E}_{VB}(x,k_y)$ is the corresponding field after AT. Here, $k_y$ denotes the spatial frequency associated with the Fourier-transformed $y$ coordinate, while $x$ remains unchanged. For the case of an LG beam, the AT results in the well known HG intensity distribution for which the order of the polynomial is related to the magnitude of the TC. In the {\bf Supplementary Material}, we present the results obtained from the AT of LG beams, both numerically and through experimental measurements, for completeness.

\subsection{Astigmatic transformation of Bessel--Gauss beams}\label{BG}

Astigmatic transformation of BG beams have previously been investigated, including the transformation produced by cylindrical lenses~\cite{AnguianoMorales2009} and by anisotropic optical media~\cite{Khonina2016Astigmatic}. More recently, cylindrical lens-induced intensity patterns have also been used to identify the TC of high-order Bessel-like beams~\cite{Yuan2022}. However, the physical origin of the characteristic TC-dependent intensity structure in finite-energy high-order BG beams has not been explicitly established. In particular, it remains unclear why the transformed intensity distribution retains a direct relation to the azimuthal phase winding despite the oscillatory radial structure of these beams. To address this question, we derive an analytical expression for the astigmatic transformation of high-order BG beams and establish the relation between the resulting intensity structure and their TC.

\begin{figure*}[ht!]
\centering    \includegraphics[width=1\linewidth,trim=0cm 0cm 0cm 0cm, clip]{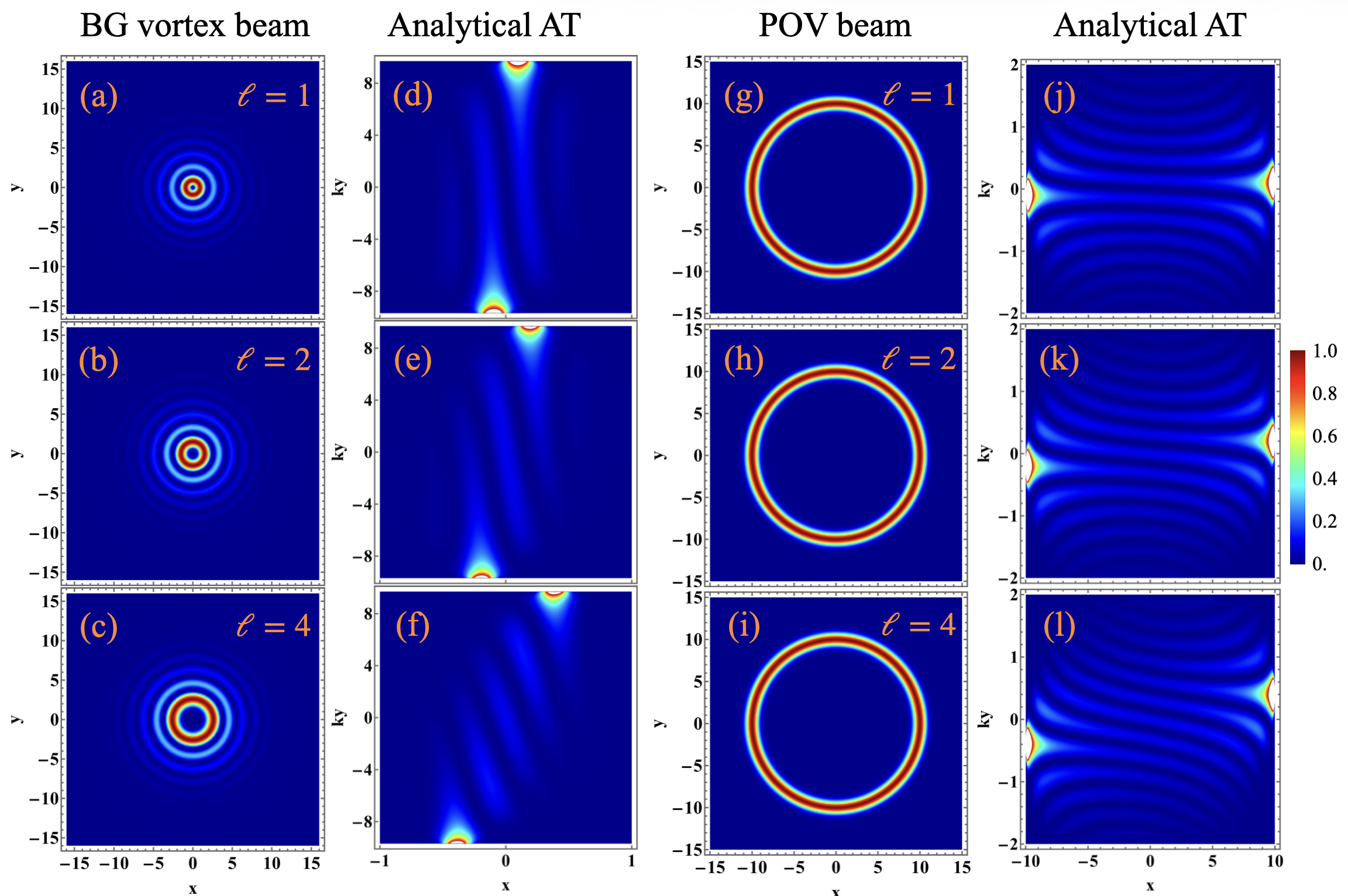}
    \caption{Theoretical transverse intensity distributions of high-order Bessel–Gaussian (BG) beams with TCs $\ell=1$, $2$, and $4$ [(a)–(c)] and their corresponding analytical ATs calculated using Eq.~(\ref{eq:BG_final}) [(d)–(f)]. Theoretical transverse intensity distributions of POV beams with $\ell=1$, $2$, and $4$ [(g)–(i)] and their corresponding analytical ATs calculated using Eq.~(\ref{Eq:ATPOV}) [(j)–(l)]. For BG beams, the number of minimum-intensity stripes in the AT profile increases with $|\ell|$, while their inclination indicates the sign of the TC.}
    \label{Fig:2}
\end{figure*}

As discussed above, high-order BG beams are finite-energy solutions of the paraxial wave equation obtained by applying a Gaussian envelope to ideal high-order Bessel beams, making them experimentally realizable. Their transverse intensity profile consists of a central dark core surrounded by concentric rings, with the radial structure primarily determined by the Bessel function of the first kind of order greater than one and with an argument that depends on the radial wave number, and modulated by the Gaussian envelope. High-order BG beams also exhibit an extended quasi-non-diffracting propagation range, over which their characteristic radial structure is approximately preserved before diffraction associated with the Gaussian envelope becomes significant. Mathematically, a high-order BG vortex beam can be represented by the following complex field amplitude:
\begin{eqnarray}
E_{BGV}(x,y) &=& E_0 J_{\ell}(k_r \sqrt{x^2+y^2})   \mathrm{e}^{-\frac{x^2+y^2}{w_{0}^2}} \nonumber \\ 
&\times&\mathrm{e}^{i\ell \arctan\left(\frac{y}{x}\right)}
\label{eqnBGsource},
\end{eqnarray}
here, $k_r=\sqrt{k^2-k_{z}^2}$, $k_{z}$, $k=2\pi/\lambda$, and $\lambda$ denote the radial wave number, longitudinal wave number, total wave number, and wavelength of the beam, respectively. Furthermore, $E_{0}$ represents the field's amplitude. After substituting Eq.~(\ref{eqnBGsource}) into Eq.~(\ref{eq:CL}) and evaluating the integral, we obtain the following expression for the transformed profile (see the {\bf Supplementary Material} for a detailed calculation):
\begin{eqnarray}
\tilde E_{BGV}(x,k_y) &\approx& \frac{2E_0}{\kappa} \mathrm{e}^{-\frac{x^2}{w_0^2}}
\mathrm{e}^{-\frac{(\ell-k_yx)^2}{(w_0\kappa)^2}}\nonumber \\
&\times& \cos(\Psi),\nonumber \\
\label{eq:BG_final}
\end{eqnarray}
with $\kappa=\sqrt{k_r^2-k_y^2}$ being the Fourier transformed variable and $\Psi = x\kappa + \ell\arcsin(k_y/k_r) - \ell\pi/2$. Equation~(\ref{eq:BG_final}) is valid under the condition that the maximum-intensity ring radius ($R$) is much larger than the beam waist ($w_{0}$), $R/w_0\gg1$, and can be described as a product of: (i) the envelope, the two Gaussian distributions, which typically constraints high-order BG beams to a finite area in the transverse plane, and (ii) the $\cos(\Psi)$ function, which determines the number of minima between the maximum-intensity lobes. The number of minima varies with the TC value in a manner similar to that observed for LG beams following the Hermite polynomials. More precisely, the argument of the cosine term is not a polynomial of order $\ell$ in a real variable, as in the LG beam case, but a phase that depends on $\ell$ through $\arcsin(k_y/k_r)$. The connection to the TC nevertheless follows from the total phase excursion accumulated over the region where the high-order BG beam is not negligible, $|k_y| \leq k_r$, set by the finite radius of the ring in $k$-space (see the {\bf Supplementary Material}). Over this range, the term of the phase that vary as the variables in the phase change is:
\begin{equation}
\ell \arcsin\Bigg[\frac{k_y}{k_r}\Bigg] \Bigg|_{-k_r}^{+k_r} = \ell \pi,
\label{eq:phase_swing}
\end{equation}
consequently, the cosine argument sweeps through a total phase of $\ell\pi$ as $k_y$ spans its allowed range. Since $\cos(\phi)$ has a zero every time $\phi$ advances by $\pi$, the number of minima of the cosine term across the full angular range of the beam is:
\begin{equation}
N_{\text{min}} \approx \frac{\ell\pi}{\pi} = \ell,
\label{eq:stripe_count}
\end{equation}
recovering the same correspondence between the number of intensity minima and the TC found for LG beams, despite the different functional form of Eq.~(\ref{eq:BG_final}). In this sense, the $\arcsin(k_y/k_r)$ term plays the same role that the polynomial order plays for LG beams after their AT via a cylindrical lens. The coefficient of $\ell$, integrated over the physical bandwidth $k_r$ of the beam, that fixes the number of zero crossings, rather than an algebraic zero-count of a real polynomial. For more details on the restriction to half of the azimuthal range, as discussed in Eq.~(\ref{eq:stripe_count}), we refer the reader to the {\bf Supplementary Material}. Figures \ref{Fig:2}~(a)-(c) show the transverse intensity distributions of high-order BG beams with TCs $\ell=1,2,$ and $4$, respectively. Similarly, Figs. \ref{Fig:2} (d)-(f) show their corresponding astigmatically transformed profiles calculated using Eq.~(\ref{eq:BG_final}). Notice that the number of minimum-intensity stripes between the maximum-intensity stripes corresponds directly to the magnitude of the TC, $|\ell|$. Furthermore, the stripe inclination reveals the sign of the TC, in the cases shown, a positive slope corresponds to positive values of $\ell$.

\subsection{Astigmatic transformation of perfect optical vortex beams}\label{POV}

The same approach can be extended to POV beams, allowing us to derive their corresponding AT equation to check if the astigmatically transformed profile carries TC information and how it differs from LG and BG beams' AT (see the {\bf Supplementary Material} for a detailed calculation). The defining feature of POV beams is a thin, bright annular intensity profile, in which both the ring radius and ring width remain independent of the TC. This makes POV beams a particularly useful test case for AT, as their radial intensity distribution remains fixed while the azimuthal phase winding varies with TC. Thus, any TC-dependent structure produced by the cylindrical lens must arise from the azimuthal phase rather than from a change in the radial intensity profile. Furthermore, the azimuthal phase term $e^{i\ell\phi}$ still carries an OAM proportional to $\ell$, preserving the vortex structure and phase singularity at the beam center despite the invariant ring geometry. The complex field amplitude of POV beams can be expressed in Cartesian coordinates as follows:
\begin{equation}
E_{POV}(x,y) = E_0 \mathrm{e}^{\frac{(\sqrt{x^2+y^2}-R)^2}{w_0^2}}\mathrm{e}^{i \ell \arctan\left(\frac{y}{x}\right)} \label{POVAsymptotic},
\end{equation}
here, $R$ is the ring radius and $w_0$ is the half-ring width of the POV beam. Note that Eq.~(\ref{POVAsymptotic}) describes POV beams in the asymptotic limit, in which the ring radius is much larger than the half-ring width i.e., $R \gg w_{0}$~\cite{Bikash:2026}. Under this condition, the spatial amplitude becomes completely independent of the TC, yielding a consistent transverse intensity profile regardless of the TC. The solution of the AT for POV beams can be obtained by substituting Eq.~(\ref{POVAsymptotic}) into Eq.~(\ref{eq:CL}): 
\begin{eqnarray}\label{eq:POV}
   \tilde{E}_{POV}(x,k_y) &\approx& E_0\sqrt{\pi} w'_0 \mathrm{e}^{-\frac{ w'^2_0 }{4} \Big(k_y+\ell \frac{x}{R^2}\Big)^2} \nonumber \\ &\times&\cos(\Phi),   
   \label{Eq:ATPOV}
\end{eqnarray}
where, $\Phi = \ell \arcsin(x/R)-k_ya-\ell \frac{\pi}{2}$. Here, we define $1/ (w'_0)^2 = a^2 /(R^2 w_0^2) $ and $a= \pm \sqrt{R^2-x^2}$. The astigmatically transformed beam in Eq.~(\ref{eq:POV}), also contains an interference term, $\cos(\Phi)$, which carries information about the beam's TC and varies in accordance with the TC value of the input POV beam. As in the case of high-order BG beams, the connection between the cosine argument and the TC follows from the total phase accumulated over the physical range of the beam, here bounded in $x$ rather than $k_y$. Since $a = \pm\sqrt{R^2 - x^2}$ is real only for $|x| \leq R$, the ring radius of the POV beam, the field amplitude is negligible outside this range. Over this physical range:
\begin{equation}
\ell \arcsin\Bigg[\frac{x}{R}\Bigg] \Bigg|_{-R}^{+R} = \ell \pi,
\label{eq:POV_phase_swing}
\end{equation}
consequently, the cosine argument sweeps through a total phase of $\ell\pi$ as $x$ spans the ring. This gives, as before:
\begin{equation}
N_{\text{min}} \approx \frac{\ell\pi}{\pi} = \ell,
\label{eq:POV_stripe_count}
\end{equation}
recovering the same correspondence between the number of intensity minima and the TC. The mechanism is thus identical to the BG vortex beam's case, with the phase winding occurring over the beam's spatial extent, set by the ring radius $R$, rather than over the angular-spectrum extent set by $k_r$. Figures \ref{Fig:2}~(g)-(i) show the theoretical transverse intensity profiles of POV beams with TCs $\ell=1,2,$ and $4$, respectively. The corresponding AT intensity profiles, calculated using Eq.~(\ref{Eq:ATPOV}), are shown in Figs.~\ref{Fig:2}~(j)-(l). The maximum-intensity stripes in this case are oriented in the orthogonal direction relative to the maximum-intensity stripes induced by the AT of BG beams. This is solely due to the Fourier conjugate nature between the BG and POV beams. Similar to the high-order BG beams, the number of minimum-intensity stripes indicates the magnitude of the TC, $|\ell|$, while their orientation provides information about the sign of the TC, which is positive for the cases shown here. Although high-order LG, BG, and POV beams all exhibit vortex characteristics, their AT profiles differ significantly. For example, BG and POV beams exhibit a distinct transformed spatial structure, characterized by more extended maximum-intensity lobes (or maximum-intensity stripes). In contrast, the AT of LG beams with a zero radial index gives rise to maximum-intensity lobes that are localized (see the {\bf Supplementary Material}). By localized, we mean that these lobes do not exhibit the extended structure observed for BG and POV beams. Moreover, the results in Fig.~\ref{Fig:2} demonstrate that there appears external structures in the FT plane, for BG and POV beams, which is absent in the case of LG beams with a zero radial index. When the AT is performed on LG beams with high radial indices, which displays multiple off-axis rings, we observe multiple intensity minima along both the horizontal and vertical directions in the resulting tilted HG patterns (see the {\bf Supplementary Material}). The TC and the radial index of the high-order LG beam can be related to the intensity minima in the horizontal and vertical directions as follows: Abs[TC] = Abs[(number of horizontal minima) - (number of vertical minima)], and radial index = min (number of horizontal minima, number of vertical minima). Here, Abs[...] denotes the absolute value. Despite having multiple rings in their transverse intensity distribution, LG beams with high radial indices do not exhibit extended maximum intensity lobes like high-order BG beams. Collectively, these results demonstrate the crucial role of the spatial amplitude of different types of vortex beams in governing the AT: (1) although LG beams with a zero radial index and POV beams exhibit a single intensity ring, their AT profiles differ, (2) although high-order BG beams and LG beams with high radial indices show multiple off-axis rings, their AT profiles differ. However, counting the number of intensity minima between the maximum-intensity lobes still provides the TC information in all cases. 


These two derivations reveal a common underlying mechanism despite the distinct functional forms of Eqs.~(\ref{eq:BG_final}) and~(\ref{eq:POV}) and represents the main finding of this work. In both cases, the AT encodes the TC through a $\cos(...)$ term whose argument accumulates a total phase of $\ell\pi$ across the physical range of the beam -- $|k_y|\leq k_r$ for the high-order BG beam, and $|x|\leq R$ for the POV beam -- regardless of whether that support is defined in the angular spectrum or in the real-space. This shared phase-winding argument yields the same correspondence, $N_{\text{min}} \approx \ell$, for both beams, indicating that the stripe-counting relation is not a peculiarity of a specific radial profile but a general consequence of the OAM-induced phase structure imprinted on the beam. This generality supports the use of the AT as a broadly applicable diagnostic for the TC of vortex beams, a conclusion we confirm experimentally in the following section. 

\section{Experimental measurements}
\subsection{Experimental setup}
\begin{figure*}[ht!]
    \centering
    \includegraphics[width=1\linewidth,trim= 0cm 0cm 0cm 0cm, clip]{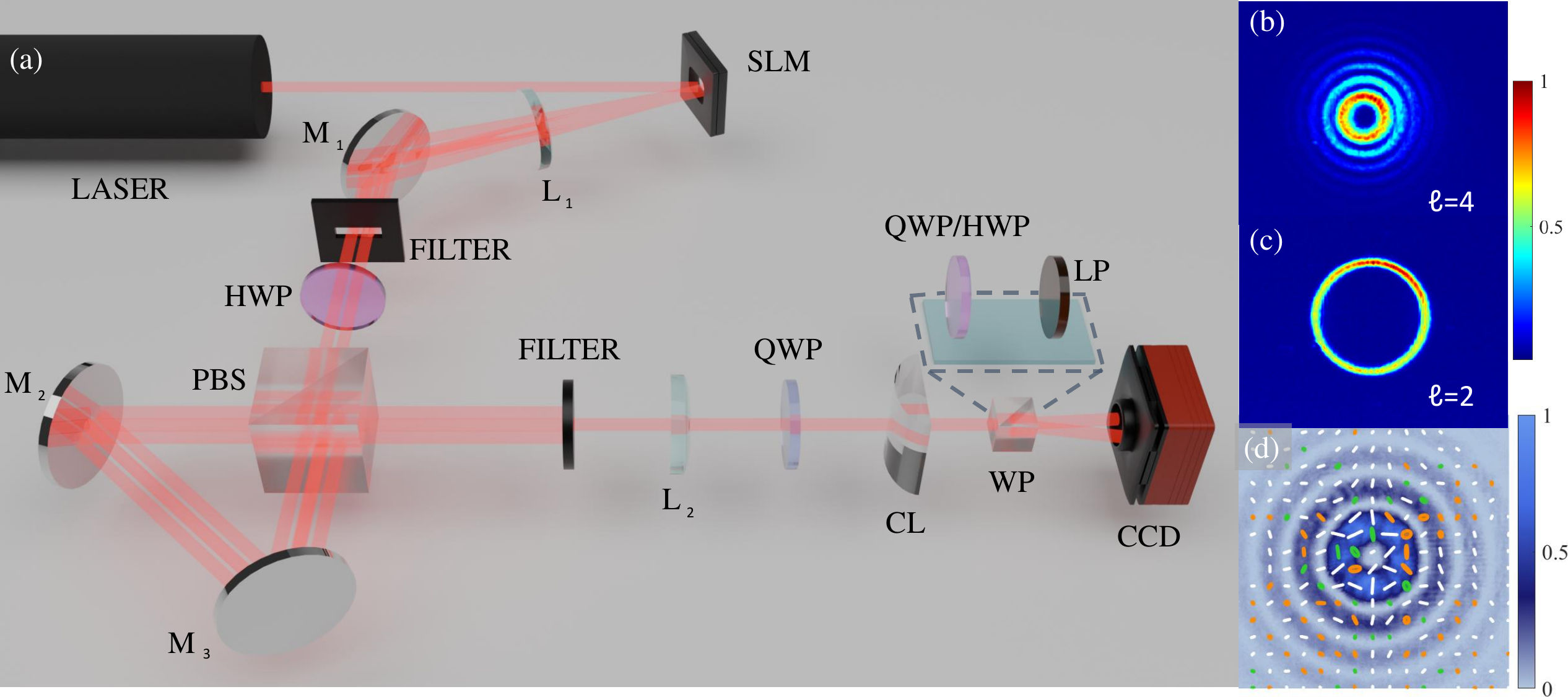}
    \caption{Schematic of the experimental setup. A collimated, linearly polarized beam illuminates a phase-only SLM displaying a multiplexed hologram. The selected diffraction orders are spatially filtered by $L_1$ and a spatial filter before entering the Sagnac interferometer, where the PBS combines orthogonally polarized modes to generate scalar or non-separable vector beams. A QWP provides polarization control, and the output is recorded by a CCD. The generated scalar BG and POV beams are shown in (b) and (c), respectively. For TC measurements, a cylindrical lens (CL) produces the characteristic intensity patterns, while a Wollaston prism (WP) separates the orthogonally polarized components of vector beams. For polarization reconstruction, the CL and WP are replaced by a linear polarizer (LP) and QWP/HWP, yielding the patterns in (d).}
    \label{Fig1ES}
\end{figure*}

The experimental setup employed for the generation and characterization of the desired scalar and vector beams is shown in Fig.~\ref{Fig1ES} (a). A collimated linearly polarized laser beam ($\lambda=633$ nm) was directed towards a reflective-type liquid-crystal spatial light modulator (SLM, Holoeye Pluto 2.1 Phase Only LCOS) with a resolution of 1920$\times$1080 and a pixel size of 8$\mu$m. The SLM displayed a single multiplexed hologram encoding the phase and amplitude information required to generate the desired scalar and vector BG beams through complex-amplitude modulation \cite{SPIEbook}. For the vector beams, the hologram simultaneously encoded two orthogonally polarized modes with TC $\ell_1$ and $\ell_2$. The desired diffraction orders were spatially isolated with a spatial filter at the focal plane of lens $L_1$. The resulting linearly polarized beam was then passed through a half-wave plate (HWP) to rotate its polarization state from horizontal to diagonal. The beam was subsequently directed towards a polarizing beam splitter (PBS), which directed the two orthogonally polarized spatial modes into a common-path interferometer (Sagnac), where they were coherently superposed to generate either scalar or vector beams. The scalar beams were obtained when only one spatial-polarization component was present, whereas simultaneous generation of both components produced a non-separable vector beam. A quarter-wave plate (QWP) at the interferometer output transforms the vector beam from the linear to the circular polarization basis. This arrangement enables the generation of arbitrary structured scalar and vector fields.

Figure \ref{Fig1ES}~(b) illustrates the intensity profile of a scalar BG beam with TC $\ell=4$, whereas Fig. \ref{Fig1ES}(c) that of a scalar POV beam with TC $\ell=2$. Similarly, Fig. \ref{Fig1ES}~(d) illustrates the intensity profile overlapped with the polarization distribution of a vector BG beam generated via the superposition of two orthogonally polarized scalar BG beams with TCs $\ell=2$ and $\ell=-2$. Furthermore, the parameters used for these beams are $w_0=2.5~\mathrm{mm}$ and $k_r=1.091\times10^{4}\mathrm{m}^{-1}$. The spatial polarization distribution of the vector beams was independently reconstructed using Stokes polarimetry. The corresponding intensity distributions were recorded with a charged-coupled device (CCD) camera by sequentially projecting the beam onto the required polarization states. This was achieved by inserting a QWP/HWP, and a linear polarizer (LP) into the beam path. Once the measurements were completed, both wave plates and the LP were removed. The measured intensities were then used to retrieve the Stokes parameters $S_0$, $S_1$, $S_2$, and $S_3$, from which the local polarization distribution was reconstructed (see the {\bf Supplementary Material} for more details).

To measure the TC, the generated beam was then transmitted through a cylindrical lens (CL), which transforms the azimuthal phase structure of the generated modes into characteristic intensity distributions. For scalar beams, the transformed intensity patterns were recorded directly with a CCD camera and subsequently used to determine the TC. For vector beams, the orthogonally polarized components were spatially separated using a Wollaston prism, allowing the TCs of the constituent modes to be determined independently from their respective intensity distributions.
\subsection{Topological charge measurements of high-order scalar Bessel-Gauss beams}

\begin{figure*}[ht!]
    \centering
    \includegraphics[width=1\linewidth,trim=0cm 0cm 0cm 0cm, clip]{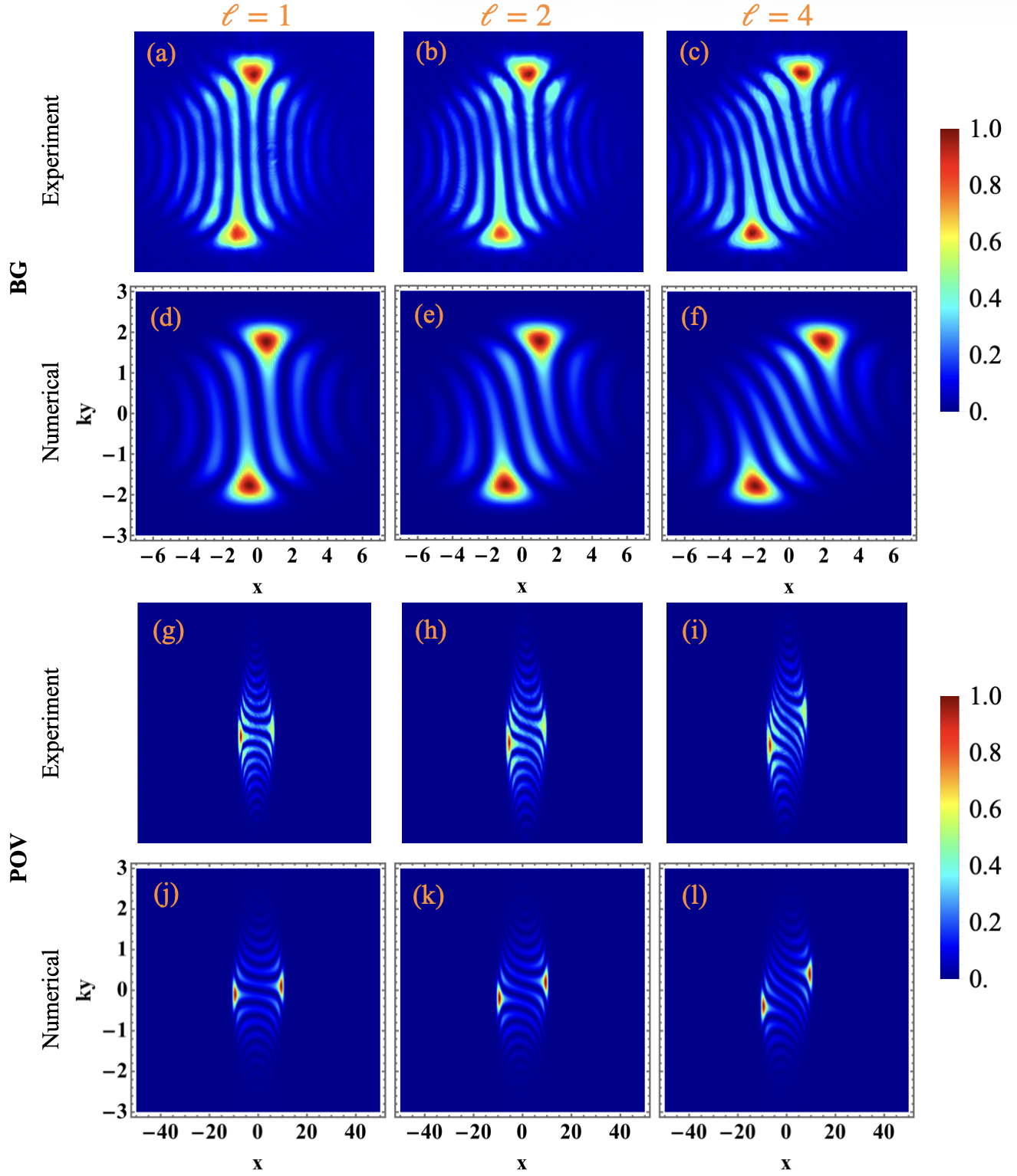}
    \caption{Experimental and numerical AT profiles for high-order BG and POV beams with TCs $\ell=1$, $2$, and $4$. The experimental results for BG beams are shown in (a)-(c), with the corresponding numerical simulation results in (d)-(f). The experimental AT profiles of the POV beams are shown in (g)-(i), together with the corresponding numerical simulation results in (j)-(l). The close agreement between the experimental and numerical simulation results demonstrates the robustness of the AT approach for identifying the TC of both high-order BG and POV beams.}
    \label{Fig3M}
\end{figure*}

To experimentally test the theoretically predicted relationship between AT and TC, we generated high-order scalar BG and POV beams with different TCs and analyzed their astigmatically transformed intensity distributions. First, we measured the astigmatically transformed profiles of the well known scalar LG beams, with additional theoretical, numerical, and experimental results presented in the {\bf Supplementary Material}. Subsequently, we generated high-order BG beams and measured their TC. The parameters used to generate BG beams from the SLM were $w_0=2.5~\mathrm{mm}$ and $k_r= 1.091\times10^{4}\mathrm{m}^{-1}$. Representative experimental measurements for BG beams with TCs $\ell=1, 2$, and $4$ are shown in Fig.~\ref{Fig3M}~(a)-(c), respectively. The corresponding AT profiles obtained from the numerical simulation are presented in Fig.~\ref{Fig3M}~(d)-(f). As demonstrated by the analytical results in Fig.~\ref{Fig:2}~(d)-(f), the number of minima occurring between the two maximum-intensity extended lobes coincides with the TC carried by the beam, imprinted on the SLM, and it is consistent with Eq.~(\ref{eq:stripe_count}). This agreement holds across the analytical, numerical, and experimental results alike. 

Furthermore, as shown in Fig.~\ref{Fig3M}~(a)-(f), the intensity distributions obtained from the numerical simulation closely resembles the experimental measurements. Both results agree well with the theoretical predictions, as shown in Fig.~\ref{Fig:2}~(d)-(f), for the TC information contained in the minima of the astigmatically transformed pattern, thereby validating the AT as a reliable approach to measure the TC of scalar high-order BG beams. We note that the intensity distributions obtained from the analytical calculation, presented in Eq.~(\ref{eq:BG_final}), is restricted to the asymptotic regime (see the {\bf Supplementary Material} for more information). A deviation from this regime yields astigmatically transformed intensity distributions that align more closely with the numerical and experimental results. Additionally, Fig.~\ref{Fig3M}~(a)-(f) reveals that both the maximum and less-intense extended lobes exhibit a more pronounced clockwise tilt as the TC increases. A similar trend is also observed for the tilted HG lobes in the case of scalar LG vortex beams, as presented in the {\bf Supplementary Material}. These results clearly demonstrate that while a cylindrical lens produces distinct transformation profiles for scalar LG and BG vortex beams, certain shared characteristics persist between the two beams.

\subsection{Topological charge measurements of scalar perfect optical vortex beams}

The theoretical results, presented in Fig.~\ref{Fig:2}~(j)-(l), confirm that the TC of scalar POV beams can be extracted in a way analogous to that used for high-order scalar LG and BG beams, validating our analytical calculation and demonstrating that the number of minimum-intensity stripes directly relates to the magnitude of the TC, while their orientation provides information about the sign of the TC. As shown in Fig.~\ref{Fig3M}~(g)-(l), for POV beams, it is clear that the number of minima between the most intense extended lobes coincides with the TC carried by the beam and imprinted on the SLM. Importantly, a comparison between LG, BG, and POV beams reveals that they exhibit distinct intensity distributions after AT. For LG beams, the HG lobes serve as the fingerprint for TC measurements. In contrast, for BG and POV beams, the lobes are extended, and the overall intensity patterns display external structures surrounding the maximum-intensity extended lobes. This underscores the structural distinctions among the beams and their imprint in the AT.  

Moreover, a direct one-to-one comparison between the AT profiles of BG and POV beams reveals that the intensity distributions in the latter are rotated by approximately $\pi/2$ relative to those in the former, a consequence arising solely from the Fourier conjugate nature of the BG and POV beams. In simple words, the Fourier transformation of BG beams produces POV beams and vice versa. However, a comparison between the AT profiles of LG and BG beams shows that the localized and extended lobes share similar tilts in the clockwise direction, when the TC is increased. However, for POV beams, the extended lobes tilt further in the counterclockwise direction as the TC increases, an effect directly opposite to that observed in high-order BG beams and reported earlier. This effect also reflects the Fourier conjugate nature of the BG and POV beams. 

Collectively, these results demonstrate that the AT provides a robust and versatile method for extracting the TC content of both scalar BG and POV beams. The excellent agreement observed between the analytical, numerical, and experimental results, for the two vortex beams with markedly different radial structures, underscores the generality of the approach, previously demonstrated for scalar LG vortex beams. In both cases, the number of minima between the maximum-intensity extended lobes of the transformed pattern serves as a direct and unambiguous signature of the TC, consistent with the prediction of Eq.~(\ref{eq:stripe_count}). The minor deviations observed in the theoretical intensity profiles are attributed to the asymptotic approximation underlying Eq.~(\ref{eq:BG_final}). These findings establish the AT as a simple, single-shot diagnostic for the TC of structured light fields, offering direct relevance to applications that demand rapid and reliable OAM characterization, as well as to spectral regimes where SLM-based techniques are not accessible -- such as for the mid-infrared region and its harmonic radiation used for the characterization of quantum materials. 

\subsection{The origin of the external structured in the AT}

Apart from whether the maximum intensity is distributed in extended or localized lobes, a key feature that further distinguishes the AT profiles of high-order BG and POV beams from those of LG vortex beams is the presence of external structures surrounding the two maximum-intensity extended lobes in the former. These external structures primarily originate from the outer rings present in scalar BG vortex beams. Interestingly, for scalar POV beams -- Fourier conjugate of scalar BG vortex beams -- similar structures are also present. 
\begin{figure}[ht!]
    \centering
    \includegraphics[width=1\linewidth,trim=0cm 0cm 0cm 0cm, clip]{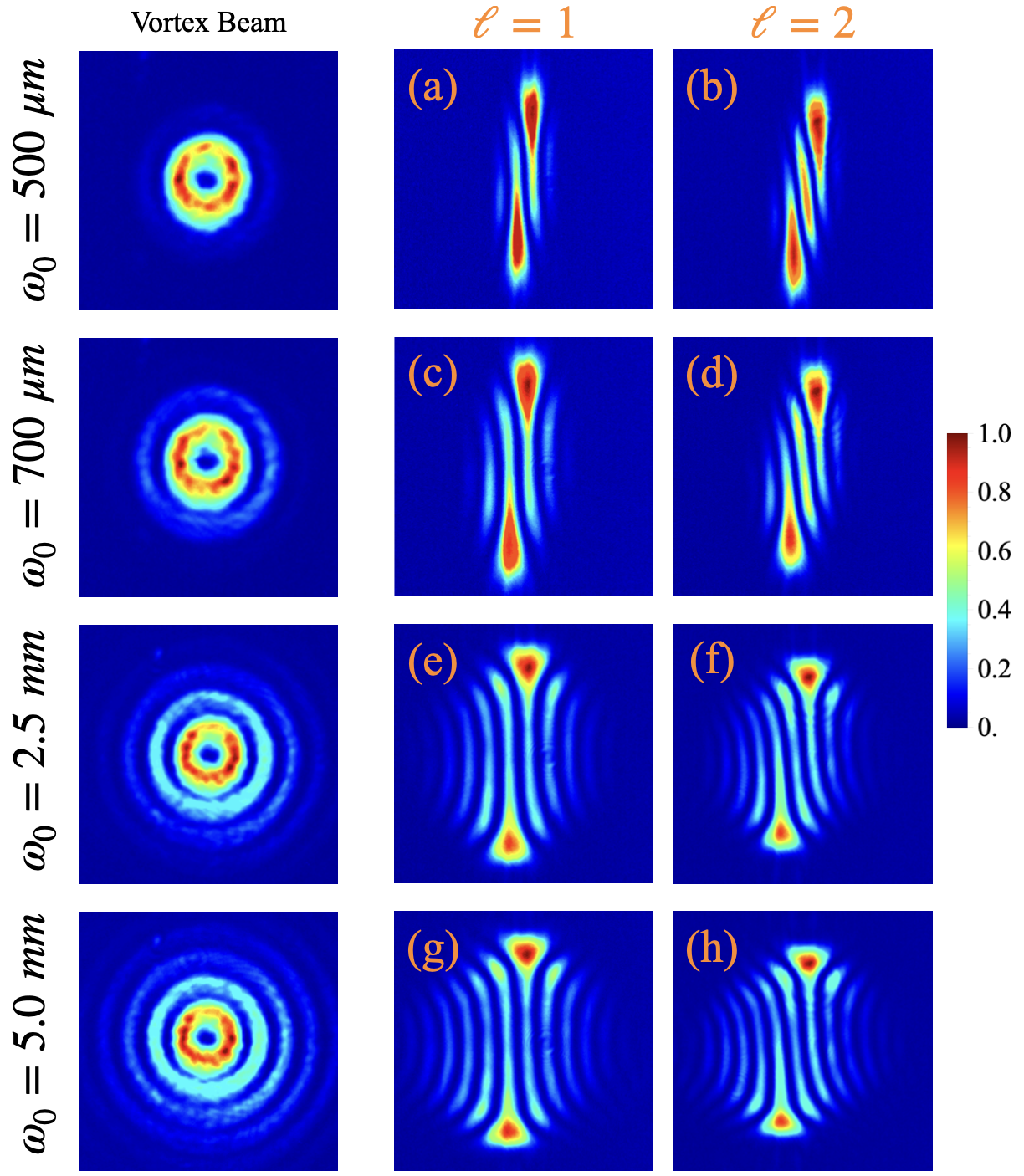}
    \caption{Role of the ring width in the AT. The beam waist $w_0$ is varied from 500~$\mu$m to 5~mm to progressively suppress the outer rings of the BGV beam. In the left column we show the resulting BGV intensity profiles for increasing $w_0$. Panels (e) to (h) show the corresponding AT for a VB with TC $\ell=1$, and panels (i) to (l) show the same for $\ell=2$. As the outer rings of the BGV are suppressed, the external structure in the AT disappears accordingly, while the central features associated with the TC remain intact.}
    \label{Fig4M}
\end{figure}
To investigate the origin of those external structures in the AT profiles of BG vortex beams, we varied the beam width and measured the corresponding transformed intensity distributions. An example of such measurements is presented Fig.~\ref{Fig4M}. The first column displays the transverse intensity distributions of the BG vortex beam for varying beam widths i.e., $w_{0}=500$~$\mu$m, $700$~$\mu$m, $2.5$~mm, and $5$~mm. It is evident from the first column that as the beam width increases, additional off-axis rings appear around the maximum-intensity ring. From the second and third columns, it is observed that a progressive increase in the beam width leads to AT profiles that exhibit more external structures surrounding the maximum-intensity extended lobes, while these external structures are suppressed at relatively smaller beam widths, as shown in panels (a), (c), (e), and (g) for $\ell=1$, and in panels (b), (d), (f), and (h) for $\ell=2$. Interestingly, the number of outer rings in scalar BG vortex beams directly relates to the number of external structures surrounding the maximum-intensity extended lobes in the transformed intensity distributions. We note that similar results were observed for scalar POV beams (see the {\bf Supplementary Material}). From a theoretical standpoint, because the maximum intensity in BG vortex beams is always concentrated on the first ring, it is sufficient to form a closed-loop that coincides with the maximum-intensity ring surrounding the central dark core in order to extract the TC information. When translated to AT profiles, this information is conveyed by the number of minima located between the maximum-intensity extended lobes. The role of additional, less-intense off-axis rings in BG vortex beams is to produce external structures surrounding the maximum-intensity extended lobes. This is precisely what we observe in the experimental results shown in Fig.~\ref{Fig4M}. Importantly, as the value of $w_0$ decreases, the multi-ring BG vortex beam evolves into a nearly-single-ring vortex beam. As a result, the corresponding AT intensity pattern closely resembles to the well-known HG-like pattern. The numerical simulation results show good agreement with the experimental findings (see the {\bf Supplementary Material}).

\section{Topological charge measurements of high-order vector Bessel-Gauss beams}

\begin{figure*}[ht!]
    \centering
    \includegraphics[width=1\linewidth,trim=0cm 0cm 0cm 0cm, clip]{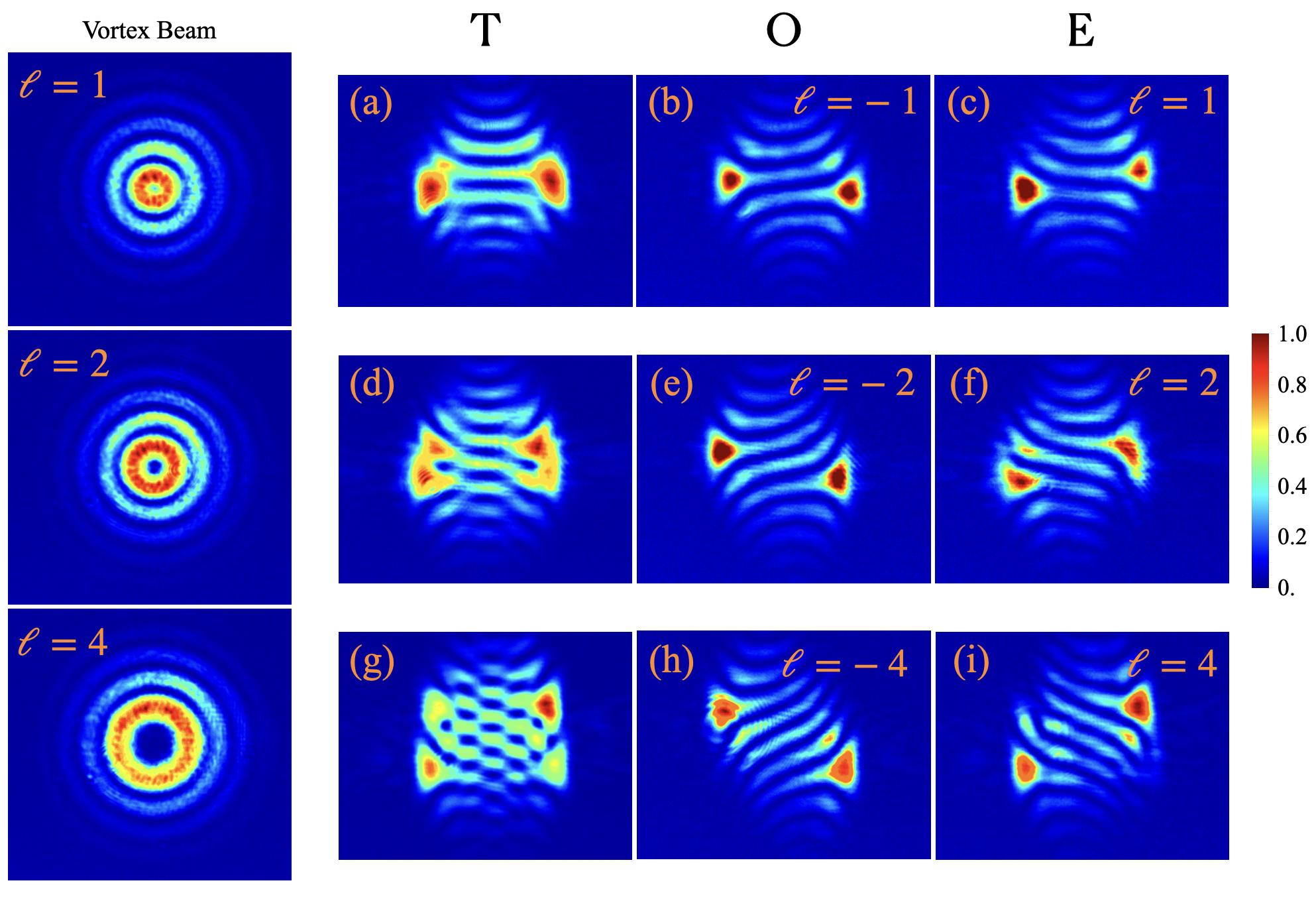}
    \caption{TC measurements of vector BG beams. The transverse intensity distributions measured for vector BG beams are shown in the leftmost panels. The measured AT profiles for the ordinary (O),  extraordinary (E), full (T) beams are presented in (a)-(i). In each case, the extraction of the TC for the different beams is unambiguous, thanks to the Wollaston prism, which enables the separation of the vector beam components and allows the minima between the maximum-intensity extended lobes -- where the TC information resides -- to be properly resolved. For every vector BG beam discussed here, the constituent scalar BG vortex beams possess the same magnitude of TC. The exact opposite orientations of the extended lobes distinguishes the positive and negative signs of the TC.}
    \label{VVBSameTC}
\end{figure*}

Building on the concept of the AT for scalar vortex beams, we extend this approach to VVBs, with a particular focus on vector BG and POV beams. Typically, vector BG (POV) beams result from the non-separable superposition of orthogonally polarized, scalar BG vortex (POV) beams. Importantly, these beam configurations provide access to a broad range of TC combinations, including: (1) symmetric pairs, with TCs of equal magnitudes and opposite signs, and (2) asymmetric pairs, characterized by TCs of distinct magnitudes and/or signs of the TCs. A representative example of the experimentally generated symmetric vector BG beams with TCs $\ell=\pm1$, $\ell=\pm2$ and $\ell=\pm4$ is shown in Fig.~\ref{VVBSameTC} (leftmost panels). The transmission of vector BG beams through the cylindrical lens produces an intricate intensity distribution that contains overlapping contributions from the two polarization components, as illustrated in Figs. \ref{VVBSameTC}~(a), (d) and (g), making it difficult to unambiguously determine the TCs of the constituent scalar BG vortex beams. However, the use of a linear polarizer allows one polarization component to be isolated at a time, enabling the TC of the constituent scalar BG vortex beams to be determined independently. Figures \ref{VVBSameTC}~(b), (e) and (h), show the TC measurements of one polarization component (the ordinary polarization, O), whereas Figs. \ref{VVBSameTC}~(c), (f) and (i) show those of the second polarization component (the extraordinary polarization, E). Importantly, the cylindrical-lens transformation itself provides a direct signature of the vectorial nature of the BG beam, as the resulting intensity structure reflects the underlying polarization-spatial coupling. In all cases, the TC extracted from the individual polarization components coincides with the TC imprinted on the SLM, confirming the validity of the AT technique for extracting the TC information from VVBs.

\begin{figure}[ht!]
    \centering
    \includegraphics[width=1\columnwidth,trim=0cm 0cm 0cm 0cm, clip]{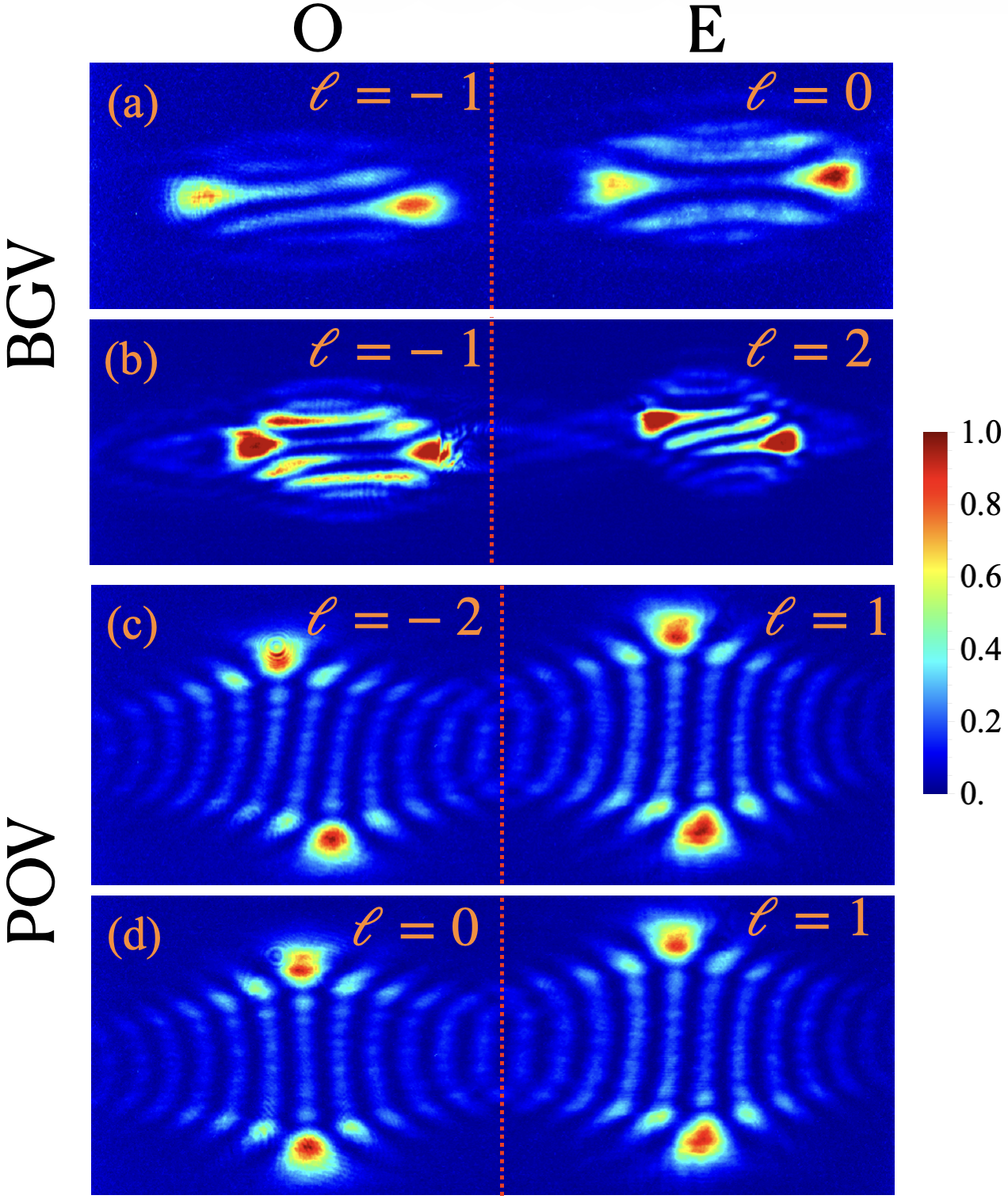}
    \caption{TC measurements of vector BG [panels (a) and (b)], and vector POV beams [panels (c) and (d)]. The extended lobes in the two beams are oriented by approximately $\pi/2$ relative to one another.} 
    \label{VVBDiffTC}
\end{figure}

 Importantly, the combination of the cylindrical lens and the Wollaston prism enables simultaneous determination of the TCs carried by each polarization component from a single measurement. Complementary measurements for vector BG beams composed of a superposition of scalar BG vortex beams with distinct TCs is presented in Fig.~\ref{VVBDiffTC}~(a)-(b). Similar to the symmetric TC pairs case, the AT profiles for different polarization components clearly reveal the TC information. Interestingly, the AT also allows for the characterization of vector BG beams with one of the components carrying a TC $\ell=0$ (i.e., zeroth-order scalar BG beam), as shown in Fig.~\ref{VVBDiffTC}~(a). Notice that for $l=0$, the maximum-intensity extended lobes perfectly overlap with each other in such a way that there exists no intensity minimum between them. The TC measurement results for vector POV beams are presented in Fig.~\ref{VVBDiffTC}~(c)-(d). Conclusions similar to vector BG beams can be drawn in this case, with the exception that the extended lobes orient by approximately $\pi/2$ relative to vector BG beams' case.   

Taken together, these results show that  vector BG beams generated by non-separable superposition of orthogonal high-order BG modes can be fully characterized using the AT. The consistency between the reconstructed polarization map and the extracted TC values validates both techniques as complementary diagnostics for vector BG beams. This dual characterization approach thus provides a complete spatial and topological description of vector BG beams within a single experimental arrangement. Most importantly, since the technique aims purely at extracting the TCs of the beams, the measurements presented here can be extended to ultrashort pulses or harmonics from the fundamental beam, where SLM-based techniques are restricted by the beam's high power and different wavelength range. 

\section{Conclusions}

Here, we demonstrated that the AT technique, traditionally employed to determine the TC of LG vortex beams, can be extended to both scalar and vector BG vortex and POV beams. Through numerical analysis of their Fourier-transformed fields and experimental validation, we showed that the TC can be reliably extracted from the resulting transformed intensity patterns in a manner analogous to that used for LG vortex beams. This experimental approach greatly facilitates the spatial characterization of the fundamental field and its high-harmonic radiation in high-power laser applications. Although the technique cannot resolve a mode superposition unless the constituent modes have comparable amplitudes, it still enables characterization of the central mode and its TC.

The ability to characterize high-power structured beams beyond conventional LG vortex modes broadens the range of vortex fields available for strong-field experiments and provides a practical route for measuring the TC of ultra-intense beams that cannot be directly generated or analyzed using conventional SLM-based methods. We anticipate that it will facilitate the implementation of advanced structured-light schemes in nonlinear optics and strong-field physics, particularly in high- harmonic generation and attosecond science.

Beyond LG, BG, and POV beams, the applicability of this mode-converter approach is not tied to a formal decomposition of the input field into HG eigenmodes, as in the case of LG vortex beams, but rather to whether the astigmatically transformed field develops an oscillatory intensity pattern whose number of lobes encodes the TC, in analogy with the node-counting rule used for HG-converted LG vortex modes. This criterion is expected to hold for other vortex-carrying fields with a smooth, singly-charged phase singularity, such as Ince--Gaussian and Mathieu--Gaussian vortex beams, where the elliptical symmetry reduces to the circular case treated here. 

The extension of the mode-converter technique to vector BG beams further broadens its applicability, since combining the AT with a Wollaston prism allows the TC of each polarization component to be extracted independently from a single measurement. This remains true both when the two components carry the same TC and when they carry different TC values, including the case where one component possesses a zero TC, indicating that the technique places no restriction on the relative OAM content of the two components. Although the AT of the full vector beam becomes increasingly blurred at higher TC due to the interference between the two constituent transformed fields, separating the polarization components with the Wollaston prism resolves this ambiguity and restores a clear TC signature for each mode. This capability is particularly relevant for the non-separable vector beams used to drive polarization-structured strong-field processes, where independent knowledge of the OAM carried by each polarization component is essential.

Apart from the TC extraction, the AT also carries information on the radial mode content of the beam. As shown in Fig.~\ref{Fig4M}, the outer rings of scalar BG and POV beams leave a corresponding signature in the number of external structures surrounding the two maximum-intensity extended lobes of the transformed pattern. This provides a simultaneous, single-shot diagnostic of both the TC and the radial mode quality of the beam, which may prove useful for monitoring the fidelity of vortex beams generated by nonlinear frequency-conversion processes.

Extending the technique to full Poincar\'e beams or to multiplexed vortex fields carrying a superposition of OAM values is also possible, and it was demonstrated here. Since TC measurements in vector beams require a mode-resolved analysis for each polarization or OAM component, such superpositions can be reliably resolved only when the constituent modes have comparable amplitudes. 

We note that the closed-form expressions derived for scalar BG and POV beams, Eqs.~(\ref{eq:BG_final}) and~(\ref{eq:POV}), respectively, are obtained under the asymptotic condition $R/w_0\gg1$. However, the underlying phase-winding argument that connects $\ell$ to the number of intensity minima, Eqs.~(\ref{eq:phase_swing})--(\ref{eq:POV_stripe_count}), does not rely on these approximations. We confirmed this directly with the experimental measurements and numerical solutions of the full 1D Fourier transform, Eq.~(\ref{eq:CL}), which reproduce the same TC-to-minima correspondence outside the asymptotic regime, establishing that the stripe-counting rule is a general property of the AT rather than an artifact of the closed-form approximations.

\section*{Acknowledgment}
C.R.G. acknowledges support from SECIHTI, México (CBF-2025-I-1804).

\section*{Disclosures}
The authors declare no conflict of interest. 

\section*{Data availability}
The data supporting the findings of this study are available from the corresponding author upon reasonable request

\section*{Supplemental document}
See Supplementary material for supporting content on the numerical simulations.

\clearpage

\bibliographystyle{unsrt}
\bibliography{References}

@article{Gori1987,
  author  = {Gori, F. and Guattari, G. and Padovani, C.},
  title   = {Bessel-Gauss beams},
  journal = {Opt. Commun.},
  volume  = {64},
  number  = {6},
  pages   = {491--495},
  year    = {1987},
  doi     = {10.1016/0030-4018(87)90276-8}
}

@article{ArltDholakia2000,
  author  = {Arlt, J. and Dholakia, K.},
  title   = {Generation of high-order Bessel beams by use of an axicon},
  journal = {Opt. Commun.},
  volume  = {177},
  number  = {1--6},
  pages   = {297--301},
  year    = {2000},
  doi     = {10.1016/S0030-4018(99)00729-1}
}

@article{Allen1992,
  author  = {Allen, L. and Beijersbergen, M. W. and Spreeuw, R. J. C. and Woerdman, J. P.},
  title   = {Orbital angular momentum of light and the transformation of Laguerre-Gaussian laser modes},
  journal = {Phys. Rev. A},
  volume  = {45},
  number  = {11},
  pages   = {8185--8189},
  year    = {1992},
  doi     = {10.1103/PhysRevA.45.8185}
}

@article{BerkhoutBeijersbergen2008,
  author  = {Berkhout, G. C. G. and Beijersbergen, M. W.},
  title   = {Method for probing the orbital angular momentum of optical vortices in electromagnetic waves from astronomical objects},
  journal = {Phys. Rev. Lett.},
  volume  = {101},
  number  = {10},
  pages   = {100801},
  year    = {2008},
  doi     = {10.1103/PhysRevLett.101.100801}
}

@article{Denisenko2009,
  author  = {Denisenko, V. and others},
  title   = {Determination of the topological charge of an optical vortex using diffraction},
  journal = {Opt. Express},
  year    = {2009}
}

@article{Liu2011,
  author  = {Liu, T. and others},
  title   = {Direct measurement of the topological charge of an optical vortex},
  journal = {Opt. Lett.},
  year    = {2011}
}

@article{Hickmann2010,
  author  = {Hickmann, J. M. and Fonseca, E. J. S. and Soares, W. C. and Chavez-Cerda, S.},
  title   = {Unveiling a truncated optical vortex lattice},
  journal = {Phys. Rev. Lett.},
  volume  = {105},
  pages   = {053904},
  year    = {2010},
  doi     = {10.1103/PhysRevLett.105.053904}
}

@article{HanZhao2011,
  author  = {Han, Y. and Zhao, G.},
  title   = {Measuring the topological charge of optical vortices with an axicon},
  journal = {Opt. Lett.},
  volume  = {36},
  number  = {11},
  pages   = {2017--2019},
  year    = {2011},
  doi     = {10.1364/OL.36.002017}
}

@article{Fu2015,
  author  = {Fu, D. and Chen, D. and Liu, R. and Wang, Y. and Gao, H. and Li, F. and Zhang, P.},
  title   = {Probing the topological charge of a vortex beam with dynamic angular double slits},
  journal = {Opt. Lett.},
  volume  = {40},
  number  = {5},
  pages   = {788--791},
  year    = {2015},
  doi     = {10.1364/OL.40.000788}
}

@article{Zhu2018,
  author  = {Zhu, J. and Zhang, P. and Chen, D. and Liu, R. and Zhou, Y. and Wang, J. and Gao, H. and Li, F.},
  title   = {Robust method to probe the topological charge of a Bessel beam by dynamic angular double slits},
  journal = {Appl. Opt.},
  volume  = {57},
  number  = {7},
  pages   = {B39--B44},
  year    = {2018},
  doi     = {10.1364/AO.57.000B39}
}

@article{Prabhakar2011,
  author  = {Prabhakar, S. and Aadhi, A. and Banerji, J. and Singh, R. P.},
  title   = {Measurement of orbital angular momentum of light using a cylindrical lens},
  journal = {Opt. Lett.},
  year    = {2011}
}

@article{Kotlyar2016,
  author  = {Kotlyar, V. V. and Kovalev, A. A. and Porfirev, A. P.},
  title   = {Determination of an optical vortex topological charge using an astigmatic transform},
  journal = {Computer Optics},
  volume  = {40},
  number  = {6},
  pages   = {781--792},
  year    = {2016},
  doi     = {10.18287/2412-6179-2016-40-6-781-792}
}

@article{Kotlyar2017,
  author  = {Kotlyar, V. V. and Kovalev, A. A. and Porfirev, A. P.},
  title   = {Astigmatic transforms of an optical vortex for measurement of its topological charge},
  journal = {Appl. Opt.},
  volume  = {56},
  number  = {14},
  pages   = {4095--4104},
  year    = {2017},
  doi     = {10.1364/AO.56.004095}
}

@article{Peng2015,
  author  = {Peng, Y. and Gan, X.-T. and Ju, P. and Wang, Y.-D. and Zhao, J.-L.},
  title   = {Measuring topological charges of optical vortices with multi-singularity using a cylindrical lens},
  journal = {Chin. Phys. Lett.},
  volume  = {32},
  number  = {2},
  pages   = {024201},
  year    = {2015},
  doi     = {10.1088/0256-307X/32/2/024201}
}

@article{Ostrovsky2013,
  author  = {Ostrovsky, A. S. and Rickenstorff-Parrao, C. and Arriz{\'o}n, V.},
  title   = {Generation of the ``perfect'' optical vortex using a liquid-crystal spatial light modulator},
  journal = {Opt. Lett.},
  volume  = {38},
  number  = {4},
  pages   = {534--536},
  year    = {2013},
  doi     = {10.1364/OL.38.000534}
}

@article{Ma2017,
  author  = {Ma, X. and J. and others},
  title   = {Perfect vortex beam: diffraction-free propagation and its applications},
  journal = {Opt. Lett.},
  volume  = {42},
  number  = {7},
  pages   = {1197--1200},
  year    = {2017}
}

@article{rosales2017simultaneous,
  title={Simultaneous generation of multiple vector beams on a single SLM},
  author={Rosales-Guzm{\'a}n, Carmelo and Bhebhe, Nkosiphile and Forbes, Andrew},
  journal={Optics express},
  volume={25},
  number={21},
  pages={25697--25706},
  year={2017},
  publisher={Optical Society of America}
}

@article{GarciaGarcia2014,
  author  = {Garc{\'\i}a-Garc{\'\i}a, J. and Rickenstorff-Parrao, C. and Ramos-Garc{\'\i}a, R. and Arriz{\'o}n, V. and Ostrovsky, A. S.},
  title   = {Simple technique for generating the perfect optical vortex},
  journal = {Opt. Lett.},
  volume  = {39},
  number  = {18},
  pages   = {5305--5308},
  year    = {2014},
  doi     = {10.1364/OL.39.005305}
}

@article{Vaity2015,
  author  = {Vaity, P. and Rusch, L.},
  title   = {Perfect vortex beam: Fourier transformation of a Bessel beam},
  journal = {Opt. Lett.},
  volume  = {40},
  number  = {4},
  pages   = {597--600},
  year    = {2015},
  doi     = {10.1364/OL.40.000597}
}

@book{SPIEbook,
Author = {Carmelo Rosales-Guzm\'{a}n and Andrew Forbes},
Publisher = {SPIE Press},
Title = {Structured light with spatial light modulators},
Year = {2024}}

@article{Padgett:96,
    author = {Padgett, M. and Arlt, J. and Simpson, N. and Allen, L.},
    title = {An experiment to observe the intensity and phase structure of Laguerre–Gaussian laser modes},
    journal = {American Journal of Physics},
    volume = {64},
    number = {1},
    pages = {77-82},
    year = {1996},
    month = {01},
    doi = {10.1119/1.18283}
}

@article{Khazaei:26,
author = {Ali Mohammad Khazaei and Saifollah Rasouli and Sergey A. Ponomarenko and Davud Hebri},
journal = {Appl. Opt.},
number = {6},
pages = {1692--1700},
title = {High-efficiency generation of Bessel--Gaussian vortex beams using Bessel-phase-embedded fork-shaped holograms},
volume = {65},
month = {Feb},
year = {2026},
doi = {10.1364/AO.581634}
}

@article{Bikash:2026,
author = {Bikash K. Das and C. Granados and M.F. Ciappina},
title = {Optical vortices: revolutionizing the field of linear and nonlinear optics},
journal = {Advances in Physics: X},
volume = {11},
number = {1},
pages = {2608076},
year = {2026},
doi = {10.1080/23746149.2025.2608076}
}

@Article{Yan:2025,
AUTHOR = {Yan, Xin and Tao, Xin and Guo, Minghao and Zhou, Chunliang and Chen, Jingzhao and Shang, Guanyu and Li, Peng},
TITLE = {Vector Vortex Beams: Theory, Generation, and Detection of Laguerre–Gaussian and Bessel–Gaussian Types},
JOURNAL = {Photonics},
VOLUME = {12},
YEAR = {2025},
NUMBER = {10},
ARTICLE-NUMBER = {1029},
DOI = {10.3390/photonics12101029}
}

@article{VAITY:2013,
title = {Measuring the topological charge of an optical vortex by using a tilted convex lens},
journal = {Physics Letters A},
volume = {377},
number = {15},
pages = {1154--1156},
year = {2013},
doi = {https://doi.org/10.1016/j.physleta.2013.02.030},
author = {Pravin Vaity and J. Banerji and R.P. Singh}
}

@article{YONGXIN:2011,
title = {Detecting the topological charge of vortex beams using an annular triangle aperture},
journal = {Optics \& Laser Technology},
volume = {43},
number = {7},
pages = {1233--1236},
year = {2011},
issn = {0030-3992},
doi = {https://doi.org/10.1016/j.optlastec.2011.03.015},
author = {Liu Yongxin and Tao Hua and Pu Jixiong and Lü Baida}
}

@Article{Cheng:2022,
AUTHOR = {Cheng, Tian-Yu and Wang, Wen-Yue and Li, Jin-Song and Guo, Ji-Xiang and Liu, Shuo and Lü, Jia-Qi},
TITLE = {Rotational Doppler Effect in Vortex Light and Its Applications for Detection of the Rotational Motion},
JOURNAL = {Photonics},
VOLUME = {9},
YEAR = {2022},
NUMBER = {7},
ARTICLE-NUMBER = {441},
DOI = {10.3390/photonics9070441}
}

@article{Panthong:18,
author = {Pituk Panthong and Sorakrai Srisuphaphon and Surasak Chiangga and Sarayut Deachapunya},
journal = {Appl. Opt.},
number = {7},
pages = {1657--1661},
title = {High-contrast optical vortex detection using the Talbot effect},
volume = {57},
month = {Mar},
year = {2018},
doi = {10.1364/AO.57.001657}
}

@article{Chen:17,
author = {Ruishan Chen and Xiaoqiang Zhang and Yong Zhou and Hai Ming and Anting Wang and Qiwen Zhan},
journal = {Appl. Opt.},
number = {16},
pages = {4868--4872},
title = {Detecting the topological charge of optical vortex beams using a sectorial screen},
volume = {56},
month = {Jun},
year = {2017},
doi = {10.1364/AO.56.004868}
}

@article{Pinnell_19,
author = {Jonathan Pinnell and Valeria Rodr\'{i}guez-Fajardo and Andrew Forbes},
journal = {Opt. Lett.},
number = {22},
pages = {5614--5617},
publisher = {Optica Publishing Group},
title = {How perfect are perfect vortex beams?},
volume = {44},
month = {Nov},
year = {2019},
doi = {10.1364/OL.44.005614}
}

@article{AnguianoMorales2009,
  author  = {Anguiano-Morales, Marcelino},
  title   = {Transformation of Bessel beams by means of a cylindrical lens},
  journal = {Applied Optics},
  year    = {2009},
  volume  = {48},
  number  = {25},
  pages   = {4826--4831},
  doi     = {10.1364/AO.48.004826}
}

@article{Khonina2016Astigmatic,
  author  = {Khonina, Svetlana N. and Paranin, Vyacheslav D. and Ustinov, Andrey V. and Krasnov, Andrey P.},
  title   = {Astigmatic transformation of Bessel beams in a uniaxial crystal},
  journal = {Optica Applicata},
  year    = {2016},
  volume  = {46},
  number  = {1},
  pages   = {5--18},
  doi     = {10.5277/oa160101}
}

@article{Yuan2022,
  author  = {Yuan, Jinpeng and Wang, Xuewen and Wang, Lirong and Xiao, Liantuan and Jia, Suotang},
  title   = {Tunable high-order Bessel-like beam generation based on cross-phase modulation},
  journal = {Optics Express},
  year    = {2022},
  volume  = {30},
  number  = {10},
  pages   = {15978--15985},
  doi     = {10.1364/OE.457232}
}

@article{Alperin2016,
author = {Samuel N. Alperin and Robert D. Niederriter and Juliet T. Gopinath and Mark E. Siemens},
journal = {Opt. Lett.},
number = {21},
pages = {5019--5022},
title = {Quantitative measurement of the orbital angular momentum of light with a single, stationary lens},
volume = {41},
month = {Nov},
year = {2016}
}

@article{Baltrukonis2021,
  title={High-order vector Bessel-Gauss beams for laser micromachining of transparent materials},
  author={Baltrukonis, Justas and Ul{\v{c}}inas, Orestas and Orlov, Sergej and Jukna, Vytautas},
  journal={Physical Review Applied},
  volume={16},
  number={3},
  pages={034001},
  year={2021},
  publisher={APS}
}

@article{yu2016super,
  title={Super-resolution deep imaging with hollow Bessel beam STED microscopy},
  author={Yu, Wentao and Ji, Ziheng and Dong, Dashan and Yang, Xusan and Xiao, Yunfeng and Gong, Qihuang and Xi, Peng and Shi, Kebin},
  journal={Laser \& Photonics Reviews},
  volume={10},
  number={1},
  pages={147--152},
  year={2016},
  publisher={Wiley Online Library}
}

@article{macdonald2002,
  title={Creation and manipulation of three-dimensional optically trapped structures},
  author={MacDonald, Michael P and Paterson, Lynn and Volke-Sepulveda, K and Arlt, Jochen and Sibbett, Wilson and Dholakia, Kishan},
  journal={Science},
  volume={296},
  number={5570},
  pages={1101--1103},
  year={2002},
  publisher={American Association for the Advancement of Science}
}

@article{fathollazade2025laguerre,
  title={Laguerre--Gaussian-to-Hermite--Gaussian mode conversion revisited},
  author={Fathollazade, Somaye and Rasouli, Saifollah and Hebri, Davud and Amiri, Pouria and Ponomarenko, Sergey A},
  journal={Journal of the Optical Society of America A},
  volume={42},
  number={4},
  pages={495--505},
  year={2025},
  publisher={Optica Publishing Group}
}

@article{fang2017examining,
  title={Examining second-harmonic generation of high-order Laguerre--Gaussian modes through a single cylindrical lens},
  author={Fang, Xinyuan and Kuang, Zeyu and Chen, Peng and Yang, Haocheng and Li, Qian and Hu, Wei and Lu, Yanqing and Zhang, Yong and Xiao, Min},
  journal={Optics Letters},
  volume={42},
  number={21},
  pages={4387--4390},
  year={2017},
  publisher={Optical Society of America}
}

@article{GranadosOAM,
  doi = {10.48550/arXiv.2601.12743},
  author = {
Debobrata Rajak and Bikash Kumar Das and Rajaram Shrestha and Bálint Kiss and Eric Cormier and Carmelo Rosales-Guzman and Stephan Fritzsche and Qiwen Zhan and Wenlong Gao and Camilo Granados},
  title = {Twisting harmonics: Transfer of orbital angular momentum in solid-state high-harmonic generation},  
  journal = {arXiv:2601.12743v1 [physics.optics] },  
  year = {2025},
  copyright = {arXiv.org perpetual, non-exclusive license}
}

@article{GranadosPOV,
  doi = {10.48550/arXiv.2512.19412},
  author = {
Camilo Granados and Bálint Kiss and Eric Cormier and Bikash Kumar Das and Debobrata Rajak and Carmelo Rosales-Guzman and Rajaram Shrestha and Qiwen Zhan and Wenlong Gao},
  title = {Sculpting ultrafast mid-infrared light for solid-state high harmonic generation},  
  journal = {arXiv:2512.19412 [physics.optics]},  
  year = {2026},
  copyright = {arXiv.org perpetual, non-exclusive license}
}

\end{document}